\documentclass[12pt,tightenlines]{revtex4}
\usepackage{amsfonts,amsmath}
\usepackage{txfonts}
\usepackage{graphicx}
\usepackage{amssymb}
\usepackage{color}
\usepackage{wrapfig,lipsum,booktabs}
\usepackage{floatrow}
\usepackage{epstopdf}
\usepackage{hyperref}
\usepackage[normalem]{ulem}

\usepackage{sidecap}

\begin{document}
\title{Topological stability of the hippocampal spatial map and synaptic transience}
\author{Yuri Dabaghian}
\affiliation{Department of Neurology\\ The University of Texas at Houston, McGovern Medical School, Houston, TX 77030, \\
$^{*}$e-mail: Yuri.A.Dabaghian@uth.tmc.edu}
\vspace{17 mm}
\date{\today}
\vspace{17 mm}
\begin{abstract}
Spatial awareness in mammals is based on internalized representations of the environment---cognitive maps---encoded 
by networks of spiking neurons. Although behavioral studies suggest that these maps can remain stable for long periods, 
it is also well-known that the underlying networks of synaptic connections constantly change their architecture due to 
various forms of neuronal plasticity. This raises a principal question: how can a dynamic network encode a stable map 
of space? In the following, we discuss some recent results obtained in this direction using an algebro-topological modeling 
approach, which demonstrate that emergence of stable cognitive maps produced by networks with transient  architectures is not
only possible, but may be a generic phenomenon. 
\end{abstract}
\maketitle

\newpage

\section{Introduction}
\label{section:intro}

\textbf{General background}. Spatial awareness in mammals is based on an internalized representation of the environment. 
Many parts of the brain are contributing to this representation, providing different types of information: cue positions 
\cite{Leathers}, geometry of the navigated paths \cite{Nitz1}, orientations \cite{Taube,MullerHD}, traveled distances 
\cite{Terrazas,Moser}, velocities \cite{Speed}, qualitative geometric relationships \cite{Nitz2,Sargolini}, etc. 
A principal question addressed by neuroscience is how all these types of data are captured by neuronal activity and what 
are the computational algorithms employed by various networks for processing this information.

\begin{figure} 
	\includegraphics[scale=0.79]{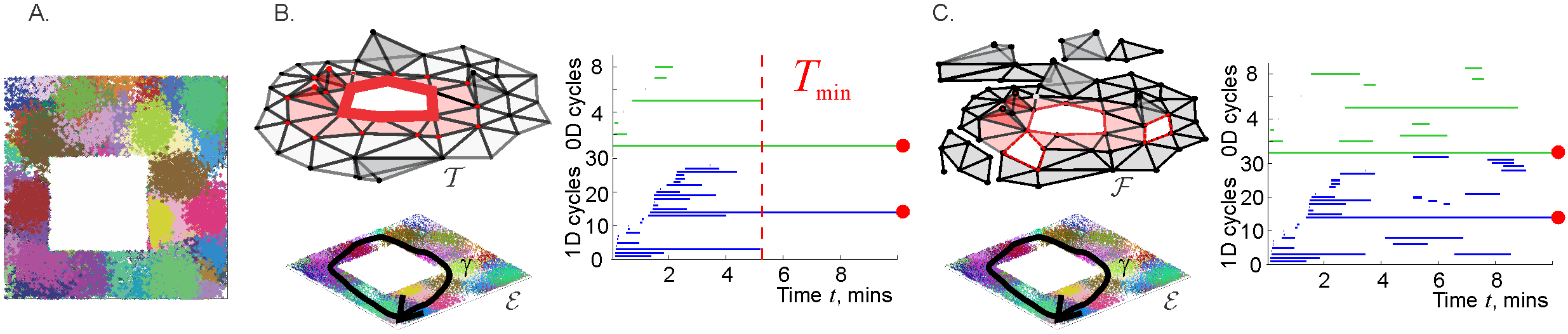}
	\caption{\label{fig:PFs} {\footnotesize
		\textbf{Place cells and cell assemblies}. (\textbf{A}). Simulated place field map $M_\mathcal{E}$ in a small 
		$(1m \times 1m)$ planar environment $\mathcal{E}$ with a square hole: dots of a particular color, marking the 
		locations where a specific place cells produced spikes, form spatial clusters---the place fields. Shown is a 
		map produced for $N=300$ place cells with a median maximal firing rate $f=14$ Hz and place field size $20$ cm. 
		(\textbf{B}). The net pool of coactivities is represented by the coactivity complex $\mathcal{T}$ (top), which 
		provides a topological representation of the environment $\mathcal{E}$ (bottom). E.g., the non-contractible 
		simplicial path shown by red chain of simplexes corresponds to a non-contractible physical path $\gamma$ around 
		the central hole in $\mathcal{E}$. The coactivity complex $\mathcal{T}$ assumes its topological shape as the 
		spatial information provided by the place cells accumulates. The panel on the right shows the timelines of 
		$0D$ (top) and $1D$ (bottom) topological loops in $\mathcal{T}$, computed using Persistent Homology theory 
		methods \cite{Carlsson1,Lum,Singh,Zomorodian}. The 
		minimal time $T_{min}$ required to eliminate the spurious loops and extract the persistent	ones (marked by 
		the red bullets) provides an estimate for the time required by a given place cell ensemble to learn the 
		topological structure of the navigated environment \cite{PLoS,Arai,Basso,Hoffman,Efficacies,Norway}.
		(\textbf{C}). If the simplexes may not only appear but also disappear, then the structure of the resulting 
		``flickering'' coactivity complex $\mathcal{F}(M_\mathcal{E})$ may never saturate, i.e., transient topological 
		defects, described by Zigzag Persistent Homology theory \cite{Carlsson2,Carlsson3,Edelsbrunner} may persist 
		indefinitely.
	}}
\end{figure}

At the current stage, our understanding of the mechanisms of spatial cognition is based mostly on empirical observations. 
For example, it was found that a major role in cognitive representation of the ambient space is played by the hippocampus: 
a vast number of experiments demonstrate that the hippocampal network contributes a ``cognitive map'' $\mathcal{C}$ that 
is crucial for the animal's ability to navigate, to find its nest and food sources, etc. \cite{Best1,OKeefe}. Experimentally, 
the properties of the cognitive map are studied by mapping hippocampal activity into the studied environment $\mathcal{E}$, 
$$f: \mathcal{C} \to \mathcal{E}.$$ In experiments with rodents (e.g., rats or mouses), this mapping is constructed by 
ascribing the ($x-y$) coordinates to very spike produced by the hippocampal principal neurons according to the animal's 
position at the time when the spike was fired \cite{SchemaS}. 
As shown in \cite{Dostrovsky}, such mapping produces spatial clusters of spikes, indicating that these neurons, known as 
the ``place cells," fire only in certain places---their respective ``place fields.'' The spatial layout of the place fields 
in $\mathcal{E}$---the place field map $M_{\mathcal{E}}$ (Fig. \ref{fig:PFs}A)---is therefore viewed as a geometric 
representation of the cognitive map of that particular environment, $\mathcal{C}(\mathcal{E})$. Electrophysiological 
recordings in ``morphing" arenas demonstrate that $M_{\mathcal{E}}$ is flexible: as the environment is slowly deformed, 
the place fields shift and change their shapes, but largely preserve their mutual overlaps, adjacency and containment 
relationships \cite{Gothard,Leutgeb,Wills,Touretzky}. Thus, the order in which the place cells spike during the animal's 
navigation remains invariant within a certain range of geometric transformations \cite{Diba,eLife,Alvernhe,Poucet,Wu,Chen}, 
which implies that $\mathcal{C}(\mathcal{E})$ may be viewed as a coarse framework of qualitative spatiotemporal relationships 
rather than precise geometry, i.e., that the hippocampal map is topological in nature.

From the computational perspective, this observation suggests that the information contained in place cell spiking should 
be interpreted topologically. In \cite{PLoS,Arai,Basso,Hoffman,Efficacies,Norway} we proposed an approach for such analyses, 
based on a schematic representations of the information supplied by place cells (co)activity. Specifically, if 
groups of coactive place cells, e.g., $c_0,c_1, \ldots,c_n$, are viewed as abstract simplexes, $\sigma=[c_0,c_1,\ldots,c_n]$, 
then the pool of the coactive place cell combinations observed by a given moment $t$ forms a simplicial ``coactivity complex'' 
$\mathcal{T}(t)$ whose topology represents the topological structure of the cognitive map of the underlying environment (see 
\cite{PLoS,Arai,Basso,Hoffman,Efficacies,Norway} and Fig.~\ref{fig:PFs}B).

The evolution of $\mathcal{T}(t)$ reflects how the net spatial information accumulates in time: starting from a few 
simplexes at the beginning of navigation, the complex $\mathcal{T}(t)$ grows and eventually, if the parameters of 
spiking activity fall within the biological range of values, assumes a shape that is topologically equivalent to the 
shape of the navigated environment in a biologically plausible period $T_{\min}$---a theoretical estimate of the time 
required to ``learn'' the environment \cite{PLoS,Arai,Basso,Hoffman,Efficacies,Norway}.

Curiously, the key building blocks of this model---the coactive groups of the hippocampal place cells represented by 
the coactivity simplexes, have physiological counterparts, called ``cell assemblies''---functionally interconnected 
groups of neurons that work as operational units of the hippocampal network \cite{Harris1,Syntax}. In \cite{CAs} it 
was shown that this correspondence can  be made accurate: the construction of the coactivity complex may be adjusted 
so that its maximal simplexes (i.e., the simplexes that are not subsimplexes of any larger simplex) represent place 
cell assemblies, rather than arbitrary combinations of coactive place cells.
An important physiological property of the cell assemblies is that these are \textit{dynamic} structures: they may form 
among the cells that demonstrate repeated coactivity and disband as a result of deterioration of synaptic connections, 
caused by reduction or cessation of spiking, then reappear due to a subsequent surge of coactivity, then disband again 
and so forth \cite{Harris1,Syntax}. In the model, the appearance and disappearance of the cell assemblies is represented 
by the the appearances and disappearances of the corresponding simplexes, so that the rewiring dynamics of the cell 
assembly network and the dynamics of the resulting cognitive map is represented by a dynamic---``flickering''---cell 
assembly complex, denoted below as $\mathcal{F}(t)$. Unlike its ``perennial" counterpart $\mathcal{T}(t)$ that can only 
grow and stabilize with time, the flickering complex $\mathcal{F}(t)$ may inflate, shrink, fragment into pieces, produce 
transient holes, fractures, gaps and other dynamic ``topological defects" (Fig.~\ref{fig:PFs}C).

Thus, on the one hand, the dynamics of $\mathcal{F}(t)$ may be viewed as a natural consequence of the network's plasticity: 
studies show that the lifetime of the ацтуал hippocampal cell assemblies ranges between minutes \cite{Billeh,Rakic,Hiratani} 
and hundreds of milliseconds \cite{Whittington,Bi}, suggesting that the hippocampal network perpetually rewires \cite{Bennett}. 
On the other hand, behavioral and cognitive studies show that spatial memories in rats can last for days and months 
\cite{Meck,Clayton,Brown2}. This poses a principal question: \textit{how can a large-scale spatial representation of 
the environment be stable if the neuronal substrate changes at a much shorter timescale}?

A principal answer to this question is suggested by an algebro-topological model of the dynamic cell assembly networks,
which allows studying the effect produced by the synaptic transience on the large-scale representation of space and 
demonstrating that a stable topological map can from within a biologically plausible period, similar to the ``perennial" 
learning period $T_{\min}(\mathcal{T})$, despite the rapid transience of the connections \cite{MWind1,MWind2,PLoZ,Replays}. 

\textbf{The large-scale topology of the cognitive map} $\mathcal{C}(\mathcal{E})$, as represented by a coactivity complex, 
can be described at different levels. A particularly concise description of a topological shape is given in terms of its 
topological loops (non-contractible surfaces identified up to topological equivalence) in different dimensions, i.e., by 
its Betti numbers $b_n$, $n=0,1,...$ \cite{Hatcher,Aleksandrov}. For example, the number of inequivalent topological loops 
that can be contracted to a zero-dimensional ($0D$) vertex, $b_0(\mathcal{F})$, corresponds to the number of the connected 
components in $\mathcal{F}(t)$; the number of loops that contract to a one-dimensional ($1D$) chain of links, 
$b_1(\mathcal{F})$, defines the number of holes and so forth \cite{Hatcher,Aleksandrov}. The full list of the Betti numbers 
of a space or a complex $X$ is known as its topological barcode, $\mathfrak{b}(X)=(b_0(X),b_1(X),b_2(X), \ldots)$, which 
captures the topological identity of $X$ \cite{Ghrist}. For example, the barcode $\mathfrak{b} = (1,1,0,\ldots)$ corresponds 
to a topological annulus, the barcode $\mathfrak{b} = (1,0,1,0, \ldots)$---to a two-dimensional ($2D$) 
sphere $S^2$, the barcode $\mathfrak{b} = (1,2,1,0, \ldots)$---to a torus $T^2$ and so forth \cite{Singh}. 
Thus, by comparing the barcode of the coactivity complex $\mathfrak{b}(\mathcal{F})$ to the barcode of the environment 
$\mathfrak{b}(\mathcal{E})$ one can establish whether their topological shapes match, 
$\mathfrak{b}(\mathcal{F}(t))= \mathfrak{b}(\mathcal{E})$, i.e., whether the coactivity complex provides a faithful 
representation of the environment at a given moment $t$. The mathematical methods required for these analyses---Persistent 
Homology \cite{Carlsson1,Lum,Singh,Zomorodian} and Zigzag Persistent Homology theories \cite{Carlsson2,Carlsson3,Edelsbrunner},
also outlined in \cite{ZomorodianBook,EdelsBook}, allow building a dynamical model of the cognitive map and addressing the
question ``\textit{How can a rapidly rewiring network produce and sustain a stable cognitive map}?"

\section{Overview of the results}
\label{section:results}

An efficient implementation of the coactivity complex $\mathcal{F}(t)$ is based on the ``cognitive graph'' model 
of the hippocampal network \cite{Muller,SchemaS}, in which each active place cell $c_i$ corresponds to a vertex 
$v_i$ of a graph $\mathcal{G}$, whose connections $\varsigma_{ij} = [v_i,v_j]$ represent pairs of coactive cells. 
An assembly of place cells $c_0,c_1, \ldots,c_n$ then corresponds to the fully interconnected subgraph, i.e., to 
a maximal clique $\varsigma=[v_0, v_1, \ldots,v_n]$ of $\mathcal{G}$ \cite{Norway,CAs,Syntax}. Since cliques, as 
combinatorial objects, can be viewed as simplexes spanned by the same sets of vertexes, the collection of 
$\mathcal{G}$-cliques defines a clique simplicial complex \cite{Jonsson} that serves as an instantiation of the 
coactivity complex \cite{Basso,Hoffman,Efficacies,Norway,CAs}. The dynamics of the clique coactivity complexes 
can be modeled based on the dynamics of the links of the corresponding coactivity graph $\mathcal{G}$. In the 
following, we discuss two such approaches, both of which demonstrate a possibility of encoding stable cognitive 
maps by transient cell assembly networks.

\subsection{Decaying clique complexes}
\label{subsection:decay}

Consider the following dynamics of the coactivity graph $\mathcal{G}$.

\begin{itemize}
	
	\item The vertexes of $\mathcal{G}$ appear when the corresponding cells become active for the first time 
	and never disappear, since according to the experiments, place cells' spiking in learned environments 
	remains stable \cite{Thompson}.
	
	\item The connection $\varsigma_{ij}$ between the vertexes $v_i$ and $v_j$ appears with probability $p_{+}=1$ 
	if the cells $c_i$ and $c_j$ become active within a $w = 1/4$ second period (for biological motivations of 
	the $w$-value see \cite{Mizuseki,Arai}). The exact time $t$ of the link's appearance can be associated with 
	any moment within $w$.
	
	\item An existing link $\varsigma_{ij}$ between cells $c_i$ and $c_j$ disappears with the probability 
	\begin{equation}
		p_{-}(t) = \frac{1}{\tau}e^{-t/\tau},
	\label{decay}
	\end{equation}
	where the time $t$ is counted from the moment of the link's last activation and $\tau$ defines the link's 
	\textit{proper} decay time. 
	
	\item The dynamics of the higher order cliques, e.g., their decay times, is fully determined by the link 
	decay period $\tau$. In the following, the notations $\mathcal{G}_{\tau}$ and $\mathcal{F}_{\tau}$ will 
	refer, respectively, to the flickering coactivity graph and the corresponding flickering clique coactivity 
	complex with the connections' proper decay rate $1/\tau$.
\end{itemize}

Note that the ongoing place cell activity can reinstate some decayed links in $\mathcal{G}_{\tau}$ and rejuvenate 
(i.e., reset the decay of) some existent ones, thus producing an \textit{effective} link's mean lifetime $\tau_{e}>\tau$ 
and leading to diverse topological dynamics of the coactivity complex $\mathcal{F}_{\tau}$. As mentioned previously, 
a key determinant of this dynamics is the sequence in which the rat traverses place fields in a map $M_{\mathcal{E}}$. 
Fixing $M_\mathcal{E}$ and the animal's trajectory $\gamma(t)$ settles the times at which place cell combinations 
become active (notwithstanding the stochasticity of neuronal firing \cite{FentonVar,Shapiro}), so that the Betti 
numbers $b_k$ of $\mathcal{F}_{\tau}(t)$ become dependent primarily on the parameters of neuronal spiking activity: 
firing rates, place field sizes, etc., and on the links' decay time $\tau$. In the following, we will review some of 
these dependencies for the case of the environment shown on Fig.~\ref{fig:PFs}A, and discuss how they affect the net 
topological structure of the corresponding cognitive map. For more details see \cite{MWind1,MWind2,PLoZ,Replays}.

\textbf{Dynamics of the decaying flickering coactivity complexes}. If $\tau$ is too small (e.g., if the coactivity 
simplexes tend to disappear between two consecutive co-activations of the corresponding cells), then the flickering 
complex should rapidly deteriorate without assuming the required topological shape. In contrast, if $\tau$ is too 
large, then the effect of the decaying connections should be small, i.e., the flickering complex $\mathcal{F}_{\tau}(t)$ 
should follow the dynamics of its ``perennial'' counterpart $\mathcal{T}(t)\equiv\mathcal{F}_{\infty}(t)$, computed 
for the same place cell spiking parameters. In particular, if the firing rates and the place field sizes are such 
that $\mathcal{T}(t)$ assumes the correct topological shape in a biologically viable time $T_{\min}(\mathcal{T})$, 
then a similar behavior should be expected from its slowly decomposing counterpart $\mathcal{F}_{\tau}(t)$. However, 
for intermediate values of $\tau$ that exceed the characteristic interval $\Delta t$ between two consecutive 
activations of a typical link in $\mathcal{G}$---a the natural timescale defined by the statistics of the rat's 
movements---the topological dynamics of $\mathcal{F}_{\tau}(t)$ may exhibit a rich variety of behaviors.

\begin{figure}[!h]
	\centering
	\includegraphics[scale=0.82]{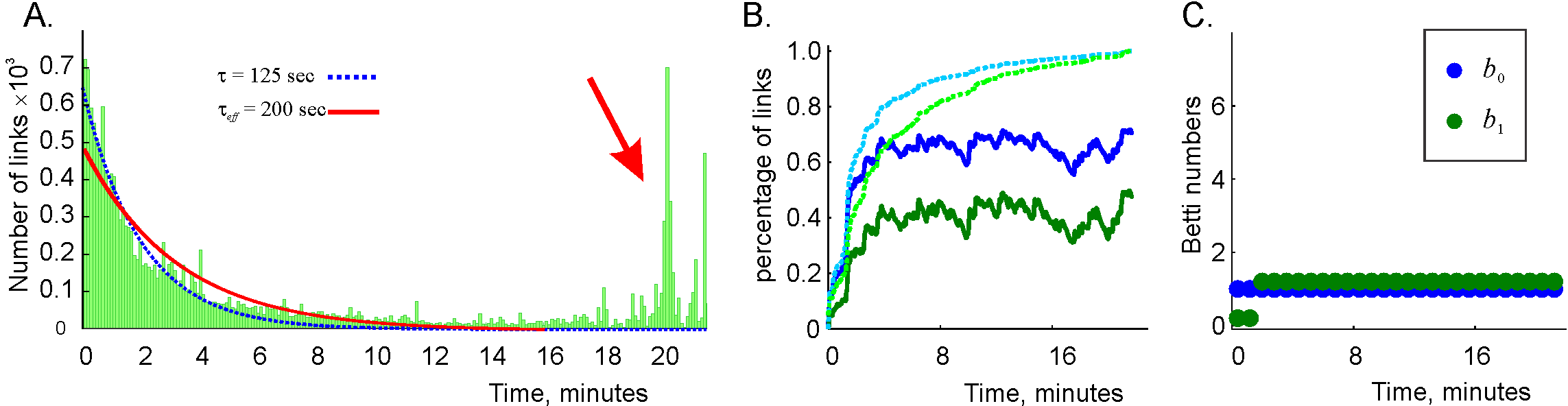}
	\caption{{\footnotesize\textbf{Topological dynamics of the decaying coactivity complex.} \textbf{A}: The histogram 
			of the connections' durations between their consecutive appearances and disappearances:	the shorter lifetimes 
			are distributed exponentially (the red line fit) and the population of the ``survivor'' links produces a bulging 
			tail of the distribution (red arrow). The dashed blue line shows the shape of the distribution (\ref{decay}). 
			\textbf{B}. The population of $1D$ (blue trace) and $2D$ (green trace) simplexes in the decaying ``flickering'' 
			complex $\mathcal{F}_{\tau}(t)$, compared to the population of $1D$ and $2D$ simplexes in the perennial complex 
			$\mathcal{T}(t)$ (dashed lines). The size of $\mathcal{F}_{\tau}(t)$ remains dynamic, whereas $\mathcal{T}(t)$ 
			saturates in about $10$ minutes.
			\textbf{C}. Betti numbers $b_0(\mathcal{F}_{\tau}(t))$ (blue) and $b_1(\mathcal{F}_{\tau}(t))$ (green) remain
			unchanged after a short initial stabilization period.}}
	\label{fig:decF}
\end{figure}

Simulations show that the characteristic inter-activation interval in the environment shown on Fig.~\ref{fig:PFs}A 
is about $\Delta t\approx 30$ seconds. For the proper decay times that generously exceed $\Delta t$, e.g., $2.5\Delta t 
\lesssim\tau\lesssim 4.5\Delta t$, the histogram of the time intervals $\Delta t_{\varsigma,i}$ between the $i^{th}$ 
consecutive birth and death of a link $\varsigma$ is bimodal: the relatively short lifetimes are exponentially distributed, with the \textit{effective} 
link lifetimes about twice higher $\tau^{(2)}_{e}\approx 2\tau$ (higher order simplexes decay more rapidly, e.g., 
$\tau^{(3)}_{e} \approx\tau$, etc.). In addition, there emerges a pool of long-living connections that persist throughout 
the entire navigation period (Fig.~\ref{fig:decF}A). In other words,  the flickering coactivity complex $\mathcal{F}_{\tau}(t)$ 
acquires a stable ``core'' formed by a population of ``surviving simplexes,'' enveloped by a population of rapidly recycling, 
``fluttering,'' simplexes. 

The numbers of $d$-dimensional simplexes in $\mathcal{F}_{\tau}(t)$ (its $f$-numbers in terminology of \cite{Gromov}) 
rapidly grow at the onset of the navigation, when $\mathcal{F}_{\tau}(t)$ inflates, but then begin to saturate by the 
time a typical link makes an appearance (in the case of the environment shown on Fig.~\ref{fig:PFs}A, this takes a few 
minutes). 
The characteristic size of $\mathcal{F}_{\tau}(t)$ grows to about a half of the size of its perennial counterpart, 
$\mathcal{F}_{\infty}(t)\equiv\mathcal{T}(t)$, with about $15\%$ fluctuations (Fig.~\ref{fig:decF}B). Thus, the population 
of simplexes in $\mathcal{F}_{\tau}(t)$ is transient: although the change of the size of $\mathcal{F}_{\tau}(t)$ 
from one moment of time to the next are relatively small, the number of simplexes that are present at a given moment 
of time $t$, but missing at a later moment $t'$, rapidly grows as a function of temporal separation $|t - t'|$, 
becoming comparable to the sizes of either $\mathcal{F}_{\tau}(t)$ or $\mathcal{F}_{\tau}(t')$ in approximately 
one effective link-decay span \cite{PLoZ,Replays}.

Meanwhile, the large-scale topology of $\mathcal{F}_{\tau}(t)$ changes significantly slower: after a brief initial 
stabilization period that roughly corresponds to the perennial learning time $T_{\min}(\mathcal{T})$, the topological 
barcode $\mathfrak{b}(\mathcal{F}_{\tau})$ remains similar to the barcode of the navigated environment $\mathcal{E}$, 
exhibiting occasional topological fluctuations at the $T_{\min}$-timescale (Fig.~\ref{fig:decF}C). Thus, the coactivity 
complex $\mathcal{F}_{\tau}$ can preserve not only its approximate size, but also its topological shape, despite the 
ongoing recycling of its simplexes.

\begin{figure}[ht] 
	\includegraphics[scale=0.84]{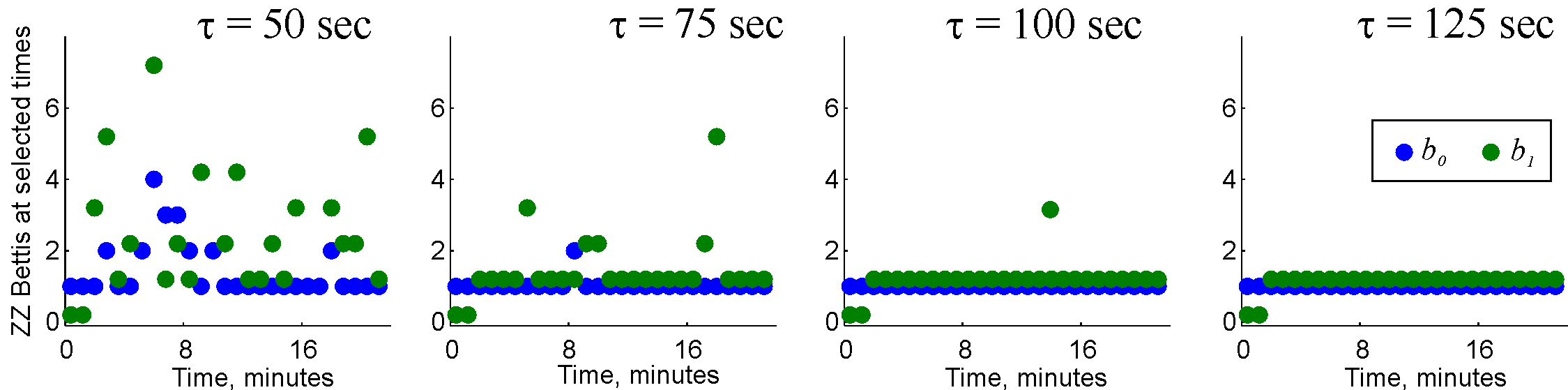}
	\caption{\label{fig:tau} {\footnotesize \textbf{Topological stabilization}. As the decay constant $\tau$ grows, the 
			topological shape of $\mathcal{F}_{\tau}(t)$ stabilizes. Shown are the Betti numbers $b_0$ (blue dots) and 
			$b_1$ (green dots) at select moments of time, computed for several values of $\tau$.
	}}
\end{figure}

As $\tau$ reduces, the topological fluctuations intensify (Fig.~\ref{fig:tau}) and vice versa, as $\tau$ grows, the 
effective lifetimes $\tau^{(2)}_{e}$ and $\tau^{(3)}_{e}$, as well as the number of the simplexes actualized at a 
given moment increase approximately linearly, resulting in a growing ``stable core'' that stabilizes the overall 
topological structure of $\mathcal{F}_{\tau}(t)$. 
Given the physiological range of parameters (simulated rat speed, place cell firing rates, place field sizes, etc.), 
a \textit{complete suppression} of topological fluctuations in the coactivity complex is achieved after the decay 
times exceed a finite threshold $\tau^{\ast}_{p}$, comparable to the time required to revisit a typical spot in the 
environment.
This value gives a theoretical estimate for the rate of physiological transience that permits stable representations 
of the environment $\mathcal{E}$ \cite{PLoZ}. 

\textbf{Alternative lifetime statistics} may strongly influence the topological dynamics of the cognitive map. 
For example, if the links' lifetimes are fixed, i.e., if the decay probability is defined by
\begin{equation}
p_{-}(t) = \begin{cases} 1 &\mbox{if } t = \tau \\ 
0 & \mbox{if } t \neq \tau,
\end{cases}
\label{fix}
\end{equation}
then the topological structure of the resulting ``quenched-decay'' coactivity complex $\mathcal{F}^{*}_{\tau}(t)$ 
changes dramatically. Even though the rejuvenation effects widen the effective distribution of the links' lifetimes 
(as before, in addition to a population of short-lived links with lifetimes close to $\tau$, there appears a 
population of the ``survivor'' simplexes), the resulting topological dynamics is more unstable: 
$\mathcal{F}^{*}_{\tau}(t)$ may split into dozens of islets containing short-lived, spurious topological defects, even 
for the values of $\tau$ that reliably produce physical Betti numbers for the exponentially distributed lifetimes 
(\ref{decay}).

As the decay slows down (i.e., as $\tau$ grows), the population of survivor links also grows and the topological 
structure of $\mathcal{F}^{*}_{\tau}(t)$ stabilizes; nevertheless, the robust, ``physical'' Betti numbers are 
attained at much (twice or more) higher values of $\tau$ than with the exponentially decaying links. 
Physiologically, this implies that the statistical spread of the connections' lifetimes (the tail of the exponential 
distribution (\ref{decay})) plays an important role: without a certain ``synaptic disorder" the network is less capable 
of capturing the topology of the environment. 

On the other hand, the topological behavior of $\mathcal{F}_{\tau}(t)$ is less sensitive to the \textit{mechanism} 
that implements a given simplex-recycling statistics. As it turns out, even if the functional connections between 
place cells are established and pruned \textit{randomly}, at a rate that matches the statistics (\ref{decay}), the 
resulting random connectivity graph $\mathcal{G}_{r}(t)$ produces a random clique complex $\mathcal{F}_{r}(t)$ with 
topological properties similar to those of $\mathcal{F}_{\tau}(t)$. In particular, the Betti numbers of $\mathcal{F}_{r}(t)$ 
converge to the Betti numbers of the environment about as quickly as the Betti numbers of its decaying counterpart 
$\mathcal{F}_{\tau}(t)$, exhibiting similar pattern of the topological fluctuations. Thus, the model suggests that 
the dynamic topology of the flickering complex may be controlled by the statistics of the decays and by the sheer 
number of simplexes present at a given moment, rather than by nature of the network's activity (e.g., random vs. 
driven by the animal's moves). 


\subsection{Finite latency complexes}
\label{subsection:intwindow}

An alternative model of flickering clique complexes can be built by restricting the period over which the 
coactivity graph is formed to a shorter time window $\varpi$ \cite{CAs}. In such approach, the coactivity 
simplexes that emerge within the starting $\varpi$-period, $\varpi_1$, will constitute a coactivity complex 
$\mathcal{F}(\varpi_1)$; the simplexes appearing within the next window, $\varpi_2$, obtained by shifting 
$\varpi_1$ over a small step $\Delta\varpi$, will form the complex $\mathcal{F}(\varpi_2)$ and so forth. 
For large consecutive window overlaps ($\Delta\varpi\ll\varpi$), a given clique-simplex $\varsigma$ (as 
defined by the set of its vertexes) may appear through a chain of consecutive windows, $\varpi_1,\varpi_2, 
\ldots, \varpi_{k-1}$, then disappear at the $k^{\textrm{th}}$ step $\varpi_k$ (i.e., $\varsigma \in 
\mathcal{F}(\varpi_{k-1}$), but $\varsigma\notin\mathcal{F}(\varpi_k)$), then reappear in a later window 
$\varpi_{l\geq k}$, then disappear again, and so forth. One may then use the midpoints $t_k$ of the windows 
in which $\varsigma$ has (re)appeared (or any other point within $\varpi_k$) to define the moments of 
$\varsigma$'s (re)births, and the matching points in the windows where it disappears to define the times 
of its deaths. By construction, the duration of $\varsigma$'s existence between its $k$-th consecutive 
appearance and disappearance, $\delta t_{\varsigma,k}$, can be as short as the shift step $\Delta\varpi$ 
or as long as the animal's navigation session. 

Simulations show that for $\varpi$ exceeding the perennial learning time $T_{min}(\mathcal{T})$ and $\Delta
\varpi\approx 0.01\varpi$, the intervals $\delta t_{\varsigma,k}$ (as well as their means averaged over $k$, 
$t_{\varsigma}= \langle\delta t_{\varsigma,k} \rangle_{k}$ and of their net existence times $\Delta T_{\varsigma}
=\sum_k \delta t_{\varsigma,k}$) are exponentially distributed, which allows characterizing the simulated 
cell assemblies by a half-life, $\tau_{\varpi}$. Specifically, for the physiological range of parameters of 
the neuronal activity in the environment shown on Fig.~\ref{fig:PFs} and $\varpi\approx 1.2 T_{\min}(\mathcal{T})$, 
the lifetime of a typical maximal simplex varies within $\tau_{\varsigma}\approx 3-12$ seconds (depending on the 
simplex' dimensionality), which is much shorter than the proper decay time in the previous model (\ref{decay}) 
and closer to the experimentally established range of values \cite{Syntax}.

\begin{figure}[ht] 
	\includegraphics[scale=0.84]{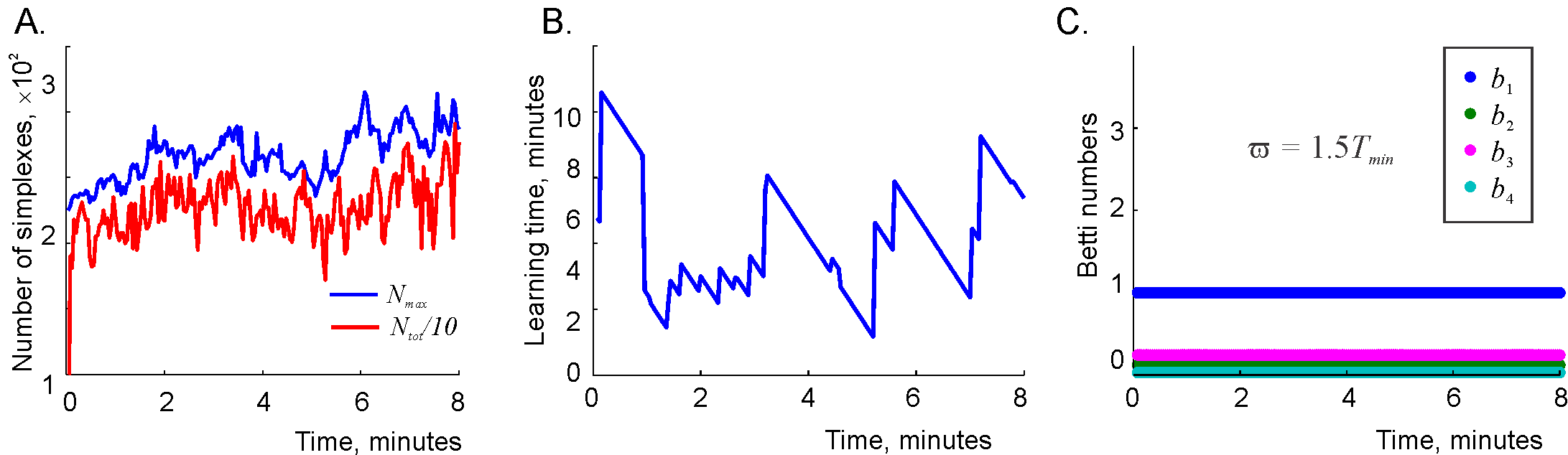}
	\caption{\label{fig:Flick} {\footnotesize \textbf{Topological dynamics in the finite latency flickering 
			complexes}. 
			\textbf{A}. The number of maximal simplexes ($N_{\max}$, blue trace) and total number of simplexes 
			($N_{tot}/10$, red trace) in the coactivity complex $\mathcal{F}_{\varpi}(t)$.
			\textbf{B}. The instantaneous learning time $T^{(k)}_{min}$ as a function of the discrete time $t_k$, 
			computed for $\varpi=1.5 T_{\min}(\mathcal{T})$.
			\textbf{C}. The low-dimensional	Betti numbers, $b_1$, $b_2$, $b_3$ and $b_4$ as a function of the 
			discrete time, computed using $\varpi=1.5 T_{\min}(\mathcal{T})$ remain stable, demonstrating full 
			topological stabilization of $\mathcal{F}_{\varpi}(t)$.
	}}
\end{figure} 

\textbf{Dynamics of the finite latency flickering coactivity complexes}. It is natural to view the individual,
``instantaneous" complexes $\mathcal{F}(\varpi_i)$ as instantiations of a single ``finite latency'' flickering 
coactivity complex, $\mathcal{F}(\varpi_i) = \mathcal{F}_{\varpi}(t_i)$. As it turns out, such complexes exhibit
a number similarities with the decaying complexes $\mathcal{F}_{\tau}(t)$. For example, the complex 
$\mathcal{F}_{\varpi}(t)$ does not fluctuate significantly: for $\varpi \geq T_{\min}(\mathcal{T})$, the number 
of simplexes contained in $\mathcal{F}_{\varpi}(t)$ changes within about $5-10\%$ of 
its mean value during the entire navigation period, but the pool of the \textit{actualized} maximal simplexes is 
renewed at about $\varpi$ timescale (Fig.~\ref{fig:Flick}A). Biologically, this implies that the simulated cell 
assembly network fully rewires over a $\varpi$ period, similar to the effective link decay time $\tau^{(2)}_{e}$ 
computed in the previous model. 

On the other hand, the large-scale topological properties of $\mathcal{F}_{\varpi}(t)$ are much more stable, 
similarly to the topological properties of $\mathcal{F}_{\tau}(t)$. For example, for sufficiently long 
latencies, $\varpi \gtrsim 1.2\, T_{\min}(\mathcal{T})$, the time required to produce the correct barcode 
$\mathfrak{b}(\mathcal{F}_{\varpi})=\mathfrak{b}(\mathcal{E})$ within each window $\varpi_k$ is typically 
finite, $T^{(k)}_{min}=T_{\min}(\mathcal{F}(\varpi_k))<\infty$ (Fig.~\ref{fig:Flick}B). Moreover, the average 
learning period $\bar{T}_{min}=\langle T^{(k)}_{min}\rangle_k$ is typically similar to the perennial learning time 
$T_{\min}(\mathcal{T})$, with a variance of about $20-40\%$ of the mean. This result shows that the topological 
dynamics in the cognitive map of a semi-randomly foraging animal is largely time-invariant, i.e., the accumulation 
of the topological information can start at any point (e.g., at the onset of the navigation) and produce the 
result in an approximately the same time period. In effect, this justifies using perennial coactivity complexes 
for estimating $T_{min}$ in \cite{PLoS,Arai,Basso,Hoffman,Efficacies,Norway}. It should also be mentioned however, 
that there also exists a number of differences between the topological dynamics of $\mathcal{F}_{\varpi}(t)$ and 
$\mathcal{F}_{\tau}(t)$, e.g., the topological fluctuations in $\mathcal{F}_{\varpi}(t)$ are mostly limited to $1D$ 
loops, $2D$ surfaces and $3D$ bubbles ($b_{0}(t)=1$, $b_{n>4}(t)=0$), whereas the fluctuations in $\mathcal{F}_{\tau}(t)$ 
also affect higher dimensions.

As $\varpi$ widens, the mean lifetime $t_{\varsigma}$ of maximal simplexes grows, suppressing the topological 
fluctuations in $\mathcal{F}_{\varpi}(t)$ and vice versa, as the memory window shrinks, the fluctuations of the 
topological loops intensify. The proportion of the ``successful'' coactivity integration windows (i.e., $\varpi_k$s 
in which the correct barcode $\mathfrak{b}(\mathcal{F}_{\varpi}(t))= \mathfrak{b}(\mathcal{E})$ is attained) 
also increases with growing $\varpi$. In fact, for $\varpi \geq\varpi_{\ast}\approx 1.5 T_{\min}$ the topological 
fluctuations tend to \textit{disappear completely} (Fig.~\ref{fig:Flick}C)---even though the simplexes' lifetimes 
remain short ($\tau^{\ast}_{\varpi}\approx 15$ secs for the environment illustrated on Fig.~\ref{fig:PFs}A).

Moreover, it can be demonstrated that as $\varpi$ exceeds a certain critical value $\varpi_{c}$ (typically 
exceeding $T_{\min}(\mathcal{T})$ by less than 40\%), the instantaneous learning times $T^{(k)}_{min}$ become 
$\varpi$-independent. Thus, the finite latency model provides a \textit{parameter-free} characterization of 
the time required by a network of place cell assemblies to represent the topology of the environment and 
establishes the timescale of the topological fluctuations in the simulated cognitive map.

\section{Discussion}

The topological model of the hippocampal cognitive map offers a connection between the spatial information 
processed by the individual place cells and the resulting global map emerging at the neuronal ensemble 
level, for both stable \cite{PLoS,Arai,Basso,Hoffman,Efficacies,Norway,CAs} and transient \cite{PLoZ,MWind1,MWind2} 
cell assembly networks. The elements of the model are embedded into the framework of simplicial topology: 
the groups of coactive cells are represented by abstract coactivity simplexes, whereas the spatial map 
encoded by the activity of neuronal populations is represented by the corresponding simplicial complexes. 
In particular, the formation and the disbanding of the cell groups is represented by the appearing and 
the disappearing coactivity simplexes, which combine into flickering coactivity complexes with nontrivial 
topological dynamics.

Generically, these dynamics occur at three principal timescales. The fastest timescale corresponds to the 
rapid recycling of the local connections---the starting point of the model. The large-scale topological 
loops, described by the time-dependent Betti numbers, unfold at a timescale that is by about an order of 
magnitude slower than the fluctuations at the simplex-level. Lastly, the topological fluctuations occur 
over certain robust base values that provide lasting, qualitative information about the environment. 

The model demonstrates that, for sufficiently slow simplex-recycling rates, the topological fluctuations at 
the intermediate timescale freeze out, i.e., the simulated cognitive map may transition into a topologically 
stable state, with static (or nearly static) Betti numbers. Physiologically, this implies that if the hippocampal 
place cell assemblies rewire sufficiently slowly, then the hippocampal map may remain stable despite the 
recycling of the connections in its neuronal substrate. Thus, the model suggests that plasticity of neuronal 
connections, which is ultimately responsible for the network's ability to incorporate new information 
\cite{McHugh,Leuner,Schaefers}, does not necessarily degrade the large-scale, qualitative information acquired 
by the system. Quite the opposite: renewing the connections allows correcting errors, e.g., removing some 
spurious, accidental topological obstructions fortuitously incorporated into the cognitive map. 
In other words, a network capable of not only accumulating, but also disposing  information, exhibits better 
learning capacity, suggesting that physiological learning should involve a balanced contribution of both 
``learning'' and ``forgetting'' components \cite{Dupret,Kuhl,Murre}.

Remarkably, the three dynamic timescales suggested by the model have their direct biological counterparts: 
the \textit{short-term memory}, which refers to temporary maintenance of ongoing (working) associations 
\cite{Cowan,Hebb}, the \textit{intermediate-term memory} that is acquired and updated at the ``operational" 
timescale \cite{Eichenbaum2,Kesner2}, and the \textit{long-term memory} that captures more persistent, 
qualitative information are broadly recognized in the literature. Physiologically, these types of memory 
are associated with different parts of the brain (hippocampal and cortical networks); thus, the model 
reaffirms functional importance of the complementary learning systems for processing spatial information 
at different levels of spatiotemporal granularity, from a theoretical viewpoint \cite{OReilly,McClelland,Fusi}.

The model allows exploring the effects produced on the cognitive map by the parameters of neuronal activity and 
the synaptic structure. For example, it can be shown, e.g., that the deterioration caused by an overly rapid 
decay  of the network's connections may be compensated by increasing neuronal activity, e.g., boosting the place 
cell firing rates \cite{PLoZ} or via contributions of the ``off-line,'' endogenous activity of the hippocampal 
network---the so-called ``place cell replays'' \cite{Karlsson1,Dragoi}. The latter are commonly viewed as 
manifestations of the animal's ``mental explorations'' of its cognitive map \cite{Johnson,Foster,Hopfield,Dabaghian} 
and are believed to help learning and to reinforce the map's stability \cite{Roux,Girardeau1}. This belief is 
largely validated by the model, which shows that sufficiently frequent, broadly distributed place cell replays, 
produced without temporal clustering, significantly reduce the topological fluctuations in the cognitive map 
$\mathcal{C}$, thus helping to separate the fast and the slow timescales and to extract stable topological 
information for the long-term, qualitative representation of the environment \cite{Replays}. Physiologically, 
these results suggest that indiscriminate, repetitive reactivations of memory sequences prevent deterioration 
of cognitive frameworks. 

Dynamical simplicial complexes have previously appeared in physical literature as discrete models of quantum 
space-time fluctuations, in the context of Simplicial Quantum Gravity theories \cite{Ambjorn,Hamber}. It was 
shown that such complexes exhibit rich geometrical and topological dynamics, e.g., they can exist in different
geometric phases, experience phase transitions between ordered and disordered states, etc., yielding regular 
behavior in the thermodynamic ``classical'' limit. 
Here dynamical complexes appear in a different context---as schematic models of the cognitive map's topological 
structure \cite{SchemaS,SchemaM}, which is naturally discrete (being encoded by finite neuronal populations) and 
transient due to the plasticity of the underlying network. 
Nevertheless, the statistical mechanics of these ``neuronal'' complexes also points at a variety of geometric and 
topological states developing at several timescales. 
In particular, using the instantaneous Betti numbers as intensive (size independent) statistical variables allows 
describing these complexes' temporal architecture and identifying the emergent topological stability phenomena.

\end{document}